**Hierarchical Relaxation in Large-twist-angle Moiré Drives Supermoiré Formation**

Chuqiao Shi[1*], Giovanny Espitia[2*], Yanxing Li[2*], Yi Jiang[3*], Yiyuan Luo[2], Xu Qi[1], Kaiwen Yang[4], Xiaohui Liu[2], Xiaoqin Li[2], Chih-Kang Shih[2#], Mit H. Naik[2,7#], Yimo Han[1,4,5,6#]

[1] Department of Materials Science and NanoEngineering, Rice University, Houston, TX, USA

[2] Department of Physics, The University of Texas at Austin, Austin, TX, USA

[3] Advanced Photon Source, Argonne National Laboratory, Lemont, IL, USA

[4] Department of Chemistry, Rice University, Houston, TX, USA

[5] Rice Advanced Materials Institute, Rice University, Houston, TX, USA

[6] Smalley-Curl Institute, Rice University, Houston, TX, USA

[7] Oden Institute for Computational Engineering and Sciences, The University of Texas at Austin, Austin, TX, USA

[*] These authors contributed equally: Chuqiao Shi, Yanxing Li, Giovanny Espitia, Yi Jiang.

[#] To whom correspondence should be addressed: Yimo Han (Email: yimo.han@rice.edu), Mit H. Naik (mit.naik@austin.utexas.edu), Chih-Kang Shih (Email: shih@mail.utexas.edu).

**Abstract**

Supermoiré lattices in large-angle twisted bilayers have recently been shown to emerge from periodic arrangements of commensurate stacking motifs, yet the microscopic structural mechanism governing their formation has remained unresolved. Here, we uncover the microscopic origin of supermoiré order as a hierarchical lattice relaxation process. Combining force-field simulations with experimental validation via super-resolution, layer-resolved multislice electron ptychography, our results identify interlayer Se–Se repulsion at the picometer scale as the driving force, giving rise to coupled short-range (~1 nm) moiré relaxation and coexisting long-range (~10 nm) supermoiré modulation. These findings demonstrate that even picometer-scale lattice relaxation remains critical in large-angle moiré systems and must be explicitly considered in understanding their emergent electronic and quantum phenomena.

## Introduction

Twisted van der Waals (vdW) bilayers provide a versatile platform for engineering emergent electronic states through moiré superlattices(*1–5*). In the small-angle regime, strong lattice relaxation reconstructs local stacking configurations, significantly modifying the electronic structure and phase behavior(*6–14*). The large moiré unit cells in these systems also enable direct probing of real-space structures using techniques such as atomic force microscopy (AFM)(*14*) , piezoresponse force microscopy (PFM)(*15*, *16*) , scanning electron microscopy (SEM)(*17*) , and dark-field transmission electron microscopy (DF-TEM)(*18*, *19*) . Quantitative measurements of lattice relaxation can be achieved using Bragg interferometry(*20–23*) and simulation-informed selected area electron diffraction(*14*, *24*, *25*) , revealing whirlpool lattice displacements on the ~20 pm scale.

By contrast, large-angle twisted bilayers have traditionally been described within a rigid-lattice approximation(*26*), where the moiré pattern is treated as a passive geometric interference(*27–30*) effect due to the relatively small lattice relaxation. However, growing experimental evidence indicates that this approximation is insufficient. Recent studies have shown that supermoiré lattices can emerge near large commensurate twist angles through periodic arrangements of distinct stacking motifs, giving rise to rich moiré induced electronic phenomena beyond the small-angle regime(*31*). Nevertheless, the microscopic structural mechanisms governing how these motifs organize into long-periodicity supermoiré order remain unresolved, particularly without explicitly accounting for lattice relaxation.

Here, we demonstrate a hierarchical lattice relaxation mechanism that drives supermoiré formation in twisted $WSe_2$ near a commensurate angle of 21.8°. Using force-field simulations, we show that this relaxation originates from interlayer Se–Se repulsion, which simultaneously induces local moiré-scale reconstruction and emergent long-periodicity supermoiré modulation. To experimentally validate this mechanism, we employ layer-resolved, super-resolution multislice electron ptychography (MEP)(*32*) , enabling direct observation of bond-length modulations arising from local moiré relaxation. Measurements of interlayer gliding and supermoiré periodicity further substantiate the presence of long-range relaxation that defines the emergent supermoiré length scale. The integration of force-field simulations with experimental validation establishes that large-angle supermoiré lattices are relaxation-driven elastic structures, rather than purely geometric interference patterns. These findings highlight that even picometer-scale atomic displacements play a decisive role in governing the structure and properties of twisted vdW systems.

**Hierarchical relaxation in large-angle moiré predicted by force-field simulation**

The macroscopic emergence of supermoiré order arises from the interplay between geometric mismatch and interlayer relaxation. We use twisted bilayer $WSe_2$ as a model system. At the exact commensurate angle of 21.8°, the bilayer forms a uniform moiré lattice with a period of 0.86 nm (**Fig. 1A**, left). At this angle, three commensurate rotational fault (CRF) configurations exist, denoted as $SE_{odd, Se}$, $SE_{odd, W}$, and $SE_{even}$, with the realized structure determined by the relative rotation center (*27*). In a purely geometric (rigid lattice) model, introducing a small twist-angle deviation (±δ) leads to additional modulation of atomic registry, producing a much longer-periodicity moiré pattern superimposed on the commensurate lattice (**Fig. 1A**, middle). The periodicity of commensurate long-periodicity modulation appears extremely sensitive to twist-angle variations (**Fig. 1B**, blue dots). However, this rigid-model prediction deviates from experimental observations(*31*) . This discrepancy indicates that the rigid lattice model fails to describe large-angle moiré systems. Interlayer relaxation must be taken into account to understand the origin of supermoiré formation (**Fig. 1A**, right).

To elucidate the microscopic mechanisms driving lattice reconstruction in large-angle moiré systems, we performed comprehensive force-field simulations. These simulations reveal that relaxation leads to a reduction in both the moiré periodicity (**Fig. 1B**, orange dots), in quantitative agreement with the experimentally observed trend in large-angle supermoiré systems. The fundamental driving force for this relaxation originates from lateral Pauli repulsion between close by non-overlapping selenium (Se) atoms(*26*, *33*), which drives opposing in-plane displacements within the two layers (**Fig. 1C**). This interaction gives rise to multiscale lattice reconstruction. At the local moiré scale (~1 nm), interlayer steric repulsion induces picometer-scale atomic displacements, resulting in distinct lattice relaxation within the three CRFs (**Fig. 1D**). This relaxation minimizes the stacking energy of each CRF (**Supplementary Table 1**), with the $SE_{odd, Se}$ configuration remaining the highest energy and thus least favorable within the supermoiré lattice.

The resulting energy imbalance between CRFs drives a secondary, long-range (~10 nm) relaxation when the twisted angle deviates slightly from the commensurate angle. This manifests as a "whirlpool"-like reconstruction centered around the high-energy $SE_{odd, Se}$ regions, where different layers have overlapped Se site in this CRF. The continuous interlayer gliding around these Se-overlapped sites introduces strong Se-Se repulsion, driving a vortex-like lattice with opposite chirality in the top and bottom layers (**Fig. 1E**). Together, these coupled processes can be summarized as a hierarchical relaxation mechanism in large-angle moiré systems (**Fig. 1F**): from atomic-scale Se–Se repulsion, to CRF-level relaxation within the moiré unit cell, and finally to long-range whirlpool reconstruction at the supermoiré scale.

**Layer-resolved multislice electron ptychography of large-angle moiré**

To experimentally verify the predicted hierarchical relaxation mechanism, we employ MEP, which provides deep-sub-ångström lateral spatial resolution and depth sensitivity(*32*) . Twisted bilayer $WSe_2$ samples were prepared at a target twist angle of $\theta \approx 21.0°$ using a tear-and-stack approach (**Method**). The twist angle was measured using nanobeam 4D-STEM, where a small convergence angle produces well-separated Bragg disks. By tracking the center of mass of these disks, we precisely determine the relative orientation between the two layers (**Fig. S1**). The sample is uniform across the field of view and exhibits a consistent supermoiré periodicity of ~9.1 nm (**Fig. S1A**). We then refined MEP to achieve super-resolution, layer-resolved imaging of the atomic structure in twisted $WSe_2$ at $\theta \approx 21.0°$ (**Fig. 2A**). This capability is enabled by optimizing both experimental conditions and ptychographic reconstruction parameters (**Supplementary Information Note.**). The approach successfully separates the overlapping atomic lattices, revealing the atomic structure of each layer across extended regions spanning multiple supermoiré periods (**Fig. 2B, C**; full images in **Fig. S2**). Comparison between the overlayed ptychographic images and a traditional annular dark field (ADF) STEM image shows strong consistency in local atomic registry, while MEP provides higher spatial resolution and explicit layer separation (**Fig. 2D**).

To quantify the confidence of layer separation in the ptychographic reconstruction, we analyze the intensity of individual atoms from two reconstructed slices of the MEP phase images. The results show that the residual contribution (defined as $I_{low}$) from the opposite layer is minimal compared to the atomic signal (defined as $I_{high}$) from the target layer (**Fig. 2E**). We define a separation degree (SD = $(I_{\text{high}} - I_{\text{low}})/(I_{\text{high}} - I_{\text{bkg}})$) to quantify this layer selectivity. Both twisted monolayer (**Fig. S3**) and twisted bilayer (**Fig. 2E**) reconstructions exhibit separation confidence exceeding 70%. Furthermore, simulations indicate that, with a sufficient electron dose (>$10^5$ e/Å$^2$), the layer-resolved MEP reconstruction achieves robust separation fidelity (**Fig. 2F,** simulation data presented in **Fig. S4-5**), supporting the reliability of the experimentally resolved atomic structures.

**Moiré relaxation in CRFs**

To provide experimental evidence for large-angle moiré relaxation, we first investigate the short-range (~1 nm) structural modulation associated with CRFs, which form the fundamental building blocks of the supermoiré structure. We performed layer resolved MEP of the three primary CRF configurations for 21.8° commensurate twisted $WSe_2$, which are distinguished by their local stacking symmetries and the relative alignment of the chalcogen and metal sublattices (**Fig. 3A-C**). These structural differences are further quantified by mapping the layer specific bond-length variations (**Fig. S6**). Bond-length measurements reveal periodic modulations characterized by consistent local "breathing" distortions (**Fig. 3D**, left; full images in **Fig. S7**), which show strong agreement with force-field predictions (**Fig. 3D**, right; full images in **Fig. S7**). Specifically, the bonds expand at the symmetry centers of the CRFs and compress toward their peripheries. These distortions originate from lateral repulsion between closely neighboring, non-overlapping chalcogen atoms in different layers, which drive them in opposite lateral directions due to Pauli repulsion (**Fig. 3E**). This localized bond-length modulation provides direct, real-space evidence of interlayer relaxation in CRFs. Furthermore, this CRF-level relaxation further modulates the relative energies of different CRF configurations (**Supplementary Table 1**), establishing the energetic landscape that governs the formation of the supermoiré structure.

**Continuous registry gliding and supermoiré relaxation**

At the long-periodicity supermoiré scale, the structure emerges from the periodic arrangement of CRFs (**Fig. 4A**). Force-field simulations show that energy differences among CRFs drive a long-range relaxation centered around the high-energy configuration, $SE_{odd, Se}$, giving rise to whirlpool-like atomic displacements (**Fig. S8**). This process is also ultimately driven by Se–Se repulsion across the interface (**Fig. S9**). This elastic self-organization redistributes local CRFs into a macroscopic periodic network, forming the supermoiré lattice, which we experimentally observe extending over domains up to ~300 nm (**Fig. S10**). However, the associated atomic displacements are less than 1 pm (**Fig. S8**), falling below the direct detection limit of MEP (*32*) .

To provide experimental evidence for supermoiré-scale relaxation, we mapped CRFs over extended regions of the lattice and found that they self-organize into a periodic superlattice with a periodicity of 9.21 nm (**Fig. 4B**). This measured period matches the predicted relaxed supermoiré lattice (**Fig. 1B**), indicating that the system undergoes the predicted whirlpool-like relaxation. To further probe the microscopic transition, we performed a gliding analysis between neighboring CRFs (**Fig. 4C-E**). We quantified the relative displacement between nearest-neighbor atoms in different layers (e.g. W and Se atoms in **Fig. 4F**), and mapped the interlayer gliding across the lateral regions connecting adjacent CRFs. The displacement evolves nearly linearly along all three principal gliding directions, with a smooth transition from overlapping atoms to laterally separated configurations of ~71 pm (**Fig. 4G**). These results indicate that, unlike small-angle moiré systems that form abrupt domain walls or rigid geometric boundaries (*14*, *15*, *34*), the lattice accommodates supermoiré strain through smooth, continuous interlayer shifts. Together, the reduced supermoiré period (9.21 nm) and the smooth, linear gliding between CRFs provide direct experimental evidence for whirlpool relaxation characterized by small, continuous atomic displacements.

## Discussion

In summary, our multiscale integration of layer-resolved ptychography and force-field simulations establishes a better understanding of large-angle twisted moiré materials at small deviations from commensurate angles. Historically, the large-angle regime has been approximated as a rigid lattice, where the emergent supermoiré order is treated as a passive, geometric optical interference effect. Our observations overturn this assumption, revealing that supermoiré lattices are actively constructed through a hierarchical elastic relaxation, fundamentally driven by inherent interlayer steric repulsion.

This hierarchical relaxation framework also clarifies a qualitative distinction from small-angle moiré systems: instead of extended domains separated by sharp domain walls, the large-angle regime supports smooth, gliding-mediated transitions that enable continuous redistribution without forming abrupt boundaries. Because competing stacking symmetries and elastic interactions are common across van der Waals heterostructures (*14*, *22*), similar constrained registry dynamics may be expected near commensurate angles in other materials systems. Structurally, such relaxation modifies local stacking environments across multiple length scales, providing a pathway for tuning electronic structure beyond simple geometric moiré engineering. In particular, the emergence of supermoiré order through coordinated gliding suggests that electronic flat bands and van Hove singularities(*31*) may be linked to this mechanism.

More broadly, the ability to directly resolve and track registry evolution opens new opportunities for studying structural dynamics in twisted quantum materials. Future work combining layer-resolved imaging with external tuning parameters such as strain, electric fields, or temperature may enable controlled manipulation of registry-state populations and provide new routes toward engineering emergent phases.

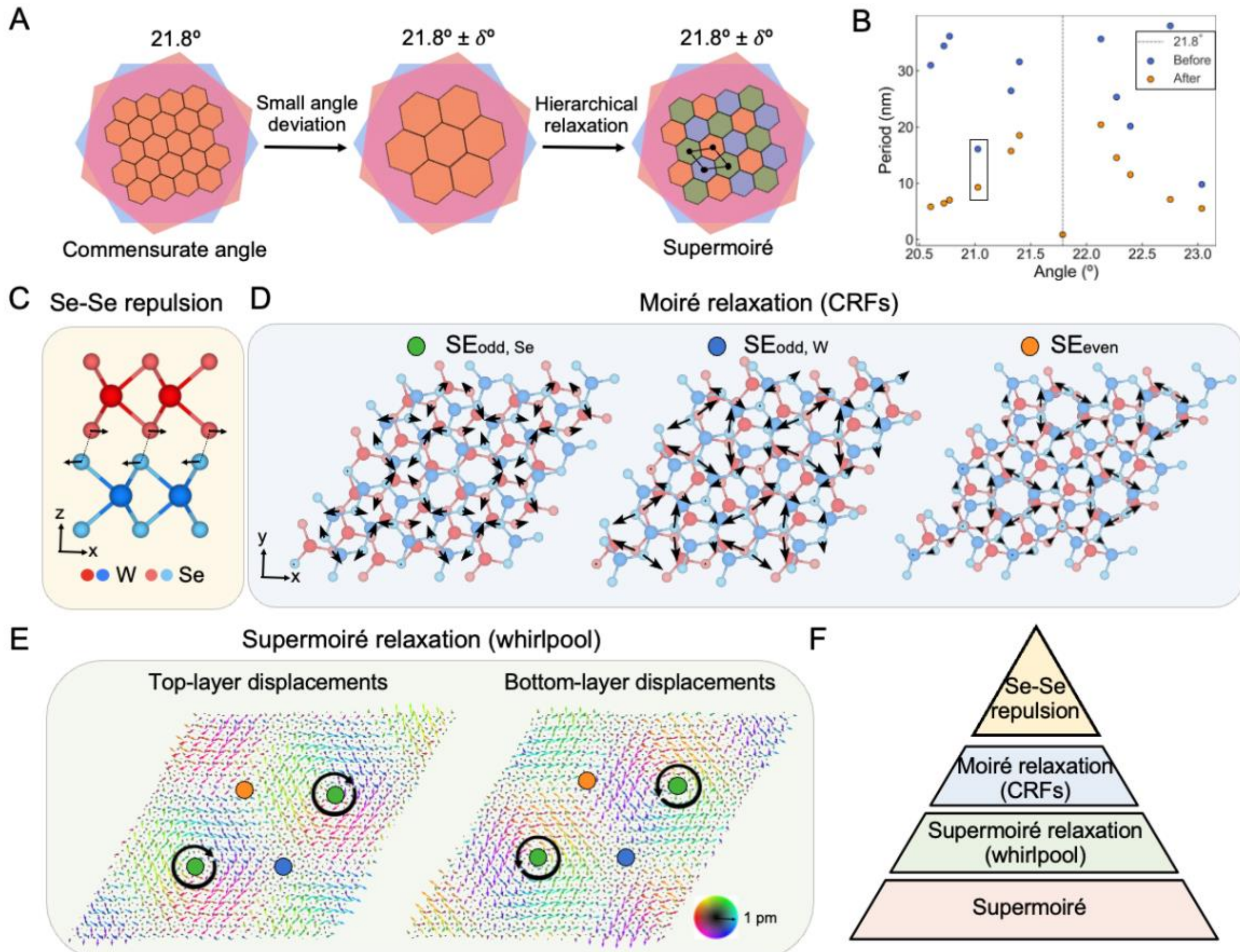

**Figure 1 | Force-field simulation of hierarchical relaxation. (A)** Schematic illustration of the origin of supermoiré formation. Left: a commensurate moiré lattice at 21.8°. Middle: long-periodicity geometric interference pattern by a slight deviation from the commensurate angle (21.8° ± δ) under the rigid-lattice approximation. Right: reduced-period supermoiré structure incorporating hierarchical relaxation. **(B)** Force-field simulations of moiré period as a function of twist-angle deviation from the commensurate angle of 21.8°. The rigid-lattice model (blue dots) consistently predicts larger moiré periods than those obtained after relaxation (orange dots) across angles near 21.8°. This study focuses on θ ≈ 21.0°, highlighted by the black box. **(C)** Schematic of interlayer Se–Se repulsion, the fundamental driving force of hierarchical relaxation. **(D)** Intralayer relaxation of the three CRFs, with black arrows indicating atomic displacement vectors driven by

Se–Se repulsion. **(E)** Interlayer "whirlpool" relaxation between distinct CRFs, driving the emergence of the supermoiré structure. **(F)** Summary of the hierarchical relaxation pyramid.

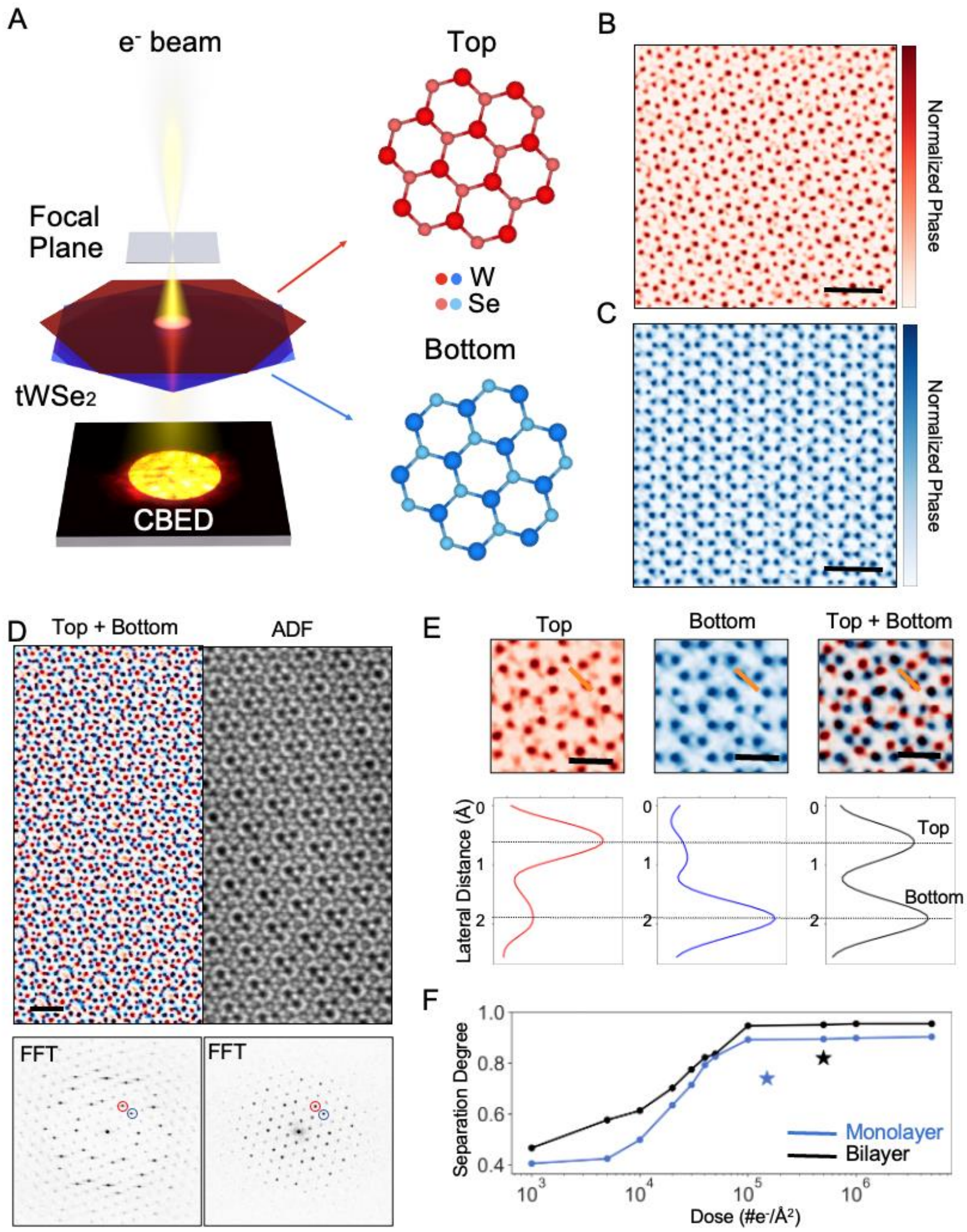

**Figure 2 | Layer-resolved multislice electron ptychography (MEP). (A)** Schematic of the MEP setup: a defocused electron probe scans the sample, and CBED patterns are recorded at each scan position. **(B, C)** Reconstructed images of the top and bottom layers of a bilayer-on-bilayer $WSe_2$ sample twisted by 21.0°. **(D)** Comparison between overlaid layer-resolved MEP images and conventional ADF-STEM images, together with the FFT, highlighting the improved spatial resolution and layer separation that enable more accurate displacement analysis. **(E)** Line profiles

of atoms from different layers, used to quantify layer separation confidence. **(F)** Separation degree extracted from MEP reconstructions of simulated data (curves) and experimental measurements (stars). Scale bar: 1nm in (B) and (C), 2 nm in (D).

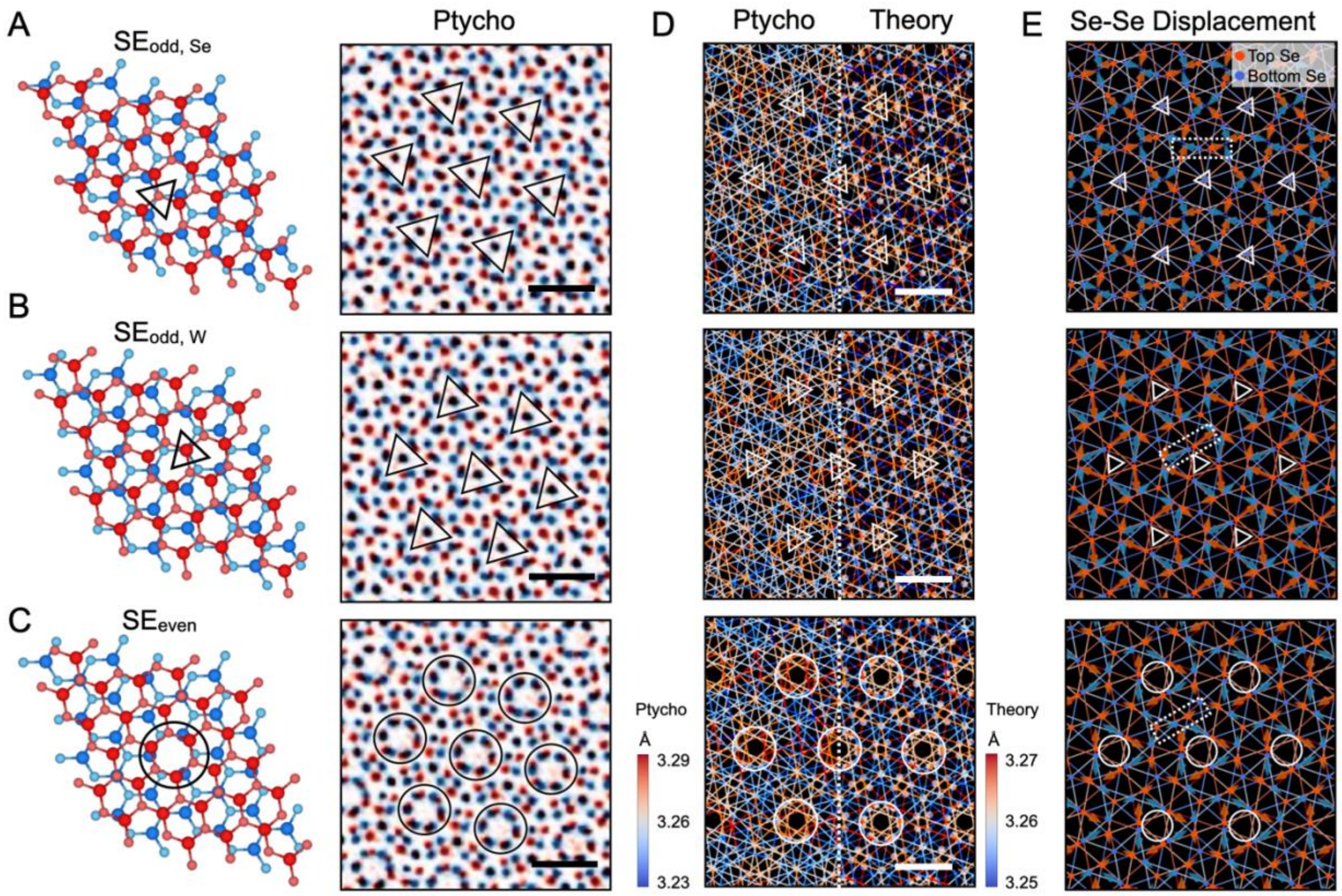

**Figure 3 | Picometer-scale lattice relaxation at CRFs. (A–C)** Atomic schematics (left) and corresponding layer-resolved ptychographic reconstructions (right) of the three primary CRF stacking configurations: (A) $SE_{odd, Se}$; (B) $SE_{odd, W}$ and (C) $SE_{even}$. Overlay markers indicate the local symmetry centers, denoted by triangles for odd configurations and circles for the even configuration. **(D)** Bond-length maps for each CRF, with split panels directly comparing experimental MEP results (left) and force-field calculations (right). **(E)** Calculated Se displacement vectors from top (red arrows) and bottom (blue arrows) layers, overlaid on force-field Se–Se bond-length maps, showing displacements in opposite directions (highlighted by white dashed boxes), indicating interlayer repulsion. Scale bars, 5 Å.

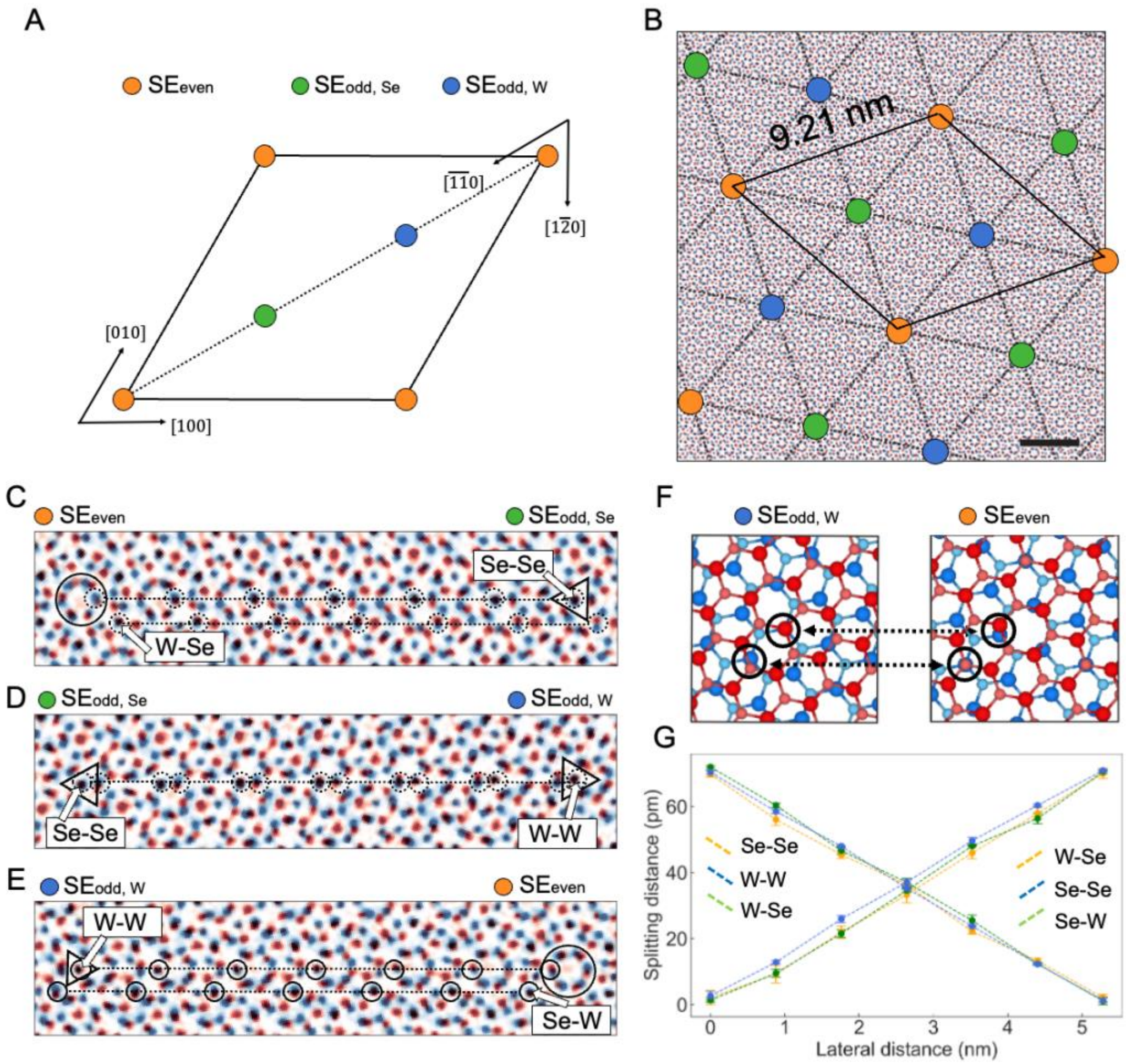

**Figure 4 | Smooth interlayer gliding in supermoiré lattices. (A)** The gliding translation diagram describing the relationships between CRFs. **(B)** Large-field-of-view MEP overlay map of the supermoiré lattice, showing a self-organized periodic network of the three distinct CRFs (indicated by colored markers) with a period of 9.21 nm. **(C-E)** MEP overlay maps of continuous atomic registry transitions between CRFs: (C) $SE_{even}$ and $SE_{odd, w}$, (D) $SE_{odd, Se}$ and $SE_{odd, w}$, and (E) $SE_{odd, w}$ and $SE_{even}$. Black (dotted) circles track the systematic lateral gliding of initially overlapping atoms from different layers. **(F)** Representative atomic model for (E), illustrating the measured displacements. Along the transition, initially overlapping W-W atoms separate laterally by ~70 pm, while previously non-overlapping Se-W atoms smoothly glide into an overlapping configuration. **(G)** Displacement profiles across the gliding transitions shown in (C–E), plotted as a function of lateral position between neighboring CRFs.

## Method

### Force Field Simulations

All commensurate superlattices were constructed using the Twister software package(*35*). Structural optimizations were carried out with classical force-field calculations as implemented in the LAMMPS(*36*) package. The Stillinger-Weber potential(*37*) was used for $WSe_2$ intralayer interactions, and Kolmogorov-Crespi(*26*) interactions were used to describe the interlayer interactions. The Kolmogorov-Crespi potential is fitted to van der Waals corrected density functional theory calculations, yielding excellent stacking fault energies and it has been used to predict structural reconstructions in quantitative agreement with experiments in several twisted bilayer transition metal dichalcogenides and their heterostructures in the past(*7*, *38–40*). All structural relaxations were performed at 0 K using the FIRE(*41*) minimization algorithm. Convergence was defined through a force tolerance of $10^{-8}$ eV/Å.

To compute the energies listed in **Supplementary Table 1**, structural optimizations and electronic structure calculations were performed using the plane-wave density functional theory (DFT) framework as implemented in the Quantum Espresso(*42*, *43*) package. Three separate structural optimizations and self-consistent field calculations were carried out, one for each commensurate rotational fault. Each structure was constructed starting from the unit cell of bilayer $WSe_2$ twisted by 21.8°. This unit cell was generated using the Twister(*35*) package, ensuring that Se atoms were aligned on top of each other in the 6-atom bilayer cell. This structure corresponds to the $SE_{odd_Se}$ configuration. The $SE_{odd_W}$ and $SE_{even}$ configurations were obtained by translating the top-layer atoms of the $SE_{odd_Se}$ by the appropriate displacement vector. Once constructed, the structures were optimized. The generalized gradient approximation was used for the exchange-correlation functional, with van der Waals interactions included via the vdw-df-c09(*44*) functional. Fully relativistic norm-conserving pseudopotentials were employed. The kinetic energy cutoff was set to 60 Ry. Atomic relaxations were performed using the Broyden-Fletcher-Goldfarb-Shanno (BFGS) algorithm with a force convergence threshold of $1 \times 10^{-3}$ atomic units (Ry/Bohr). The convergence threshold for self-consistency in the energy was set to $1.0 \times 10^{-6}$ Ry.

### Twisted $WSe_2$ sample fabrication

The free-standing TEM samples were prepared with tear-and-stack methods and Polypropylene carbon (PPC) based dry transfer technique. All $WSe_2$ layers were mechanically exfoliated from high-quality crystals by Polydimethylsiloxane (PDMS) on 285 nm $SiO_2$/Si substrates. A PPC/PDMS dome stamp is used to dry transfer the sample from a Si substrate to a TEM grid. Under a temperature of around 40°C, we tore part of the $WSe_2$ monolayer and twisted the rest part with certain angles and finally released it on a SiNx TEM grid (Norcada) at 150°C. The PPC was slowly annealed away under 250 °C at ultra-high vacuum (UHV) ($1 \times 10^{-9}$ torr) for 72 hours.

**4D-STEM data acquisition and processing**

The 4D-STEM datasets were collected using an aberration-corrected ThermoFisher Titan Themis equipped with an Electron Microscope Pixel Array Detector (EMPAD)(*45*) at an acceleration voltage of 80 keV. The EMPAD features a 128 × 128 pixel array, a 1 ms acquisition time, a readout speed of 0.86 ms per frame, and a dynamic range of 1,000,000:1. The scan step size is 0.83Å in real space. The convergence angle is 30 mrad and the camera length is 360 mm, which gives a center disk with 29-pixel radius in each diffraction pattern. The defocus is set as -20 nm.

Both the fold-slice(*46*) written in MATLAB (https://github.com/yijiang1/fold_slice) and Ptychi library written in Python (https://github.com/AdvancedPhotonSource/pty-chi/tree/pear_aps) are used for ptychographic reconstruction Supplementary Table 1 summarizes the reconstruction parameters, which influence both the quality and efficiency of ptychographic reconstruction and are frequently adjusted for different experimental data. The core algorithm is the maximum likelihood ptychography with a least-squares solver(*47*) , which provides both Gaussian and Poisson probability distribution to model data noise. Here we use the Poisson probability distribution. The method also used a mini-batch update strategy to efficiently balance reconstruction speed and convergence rate. Thus, the number of diffraction patterns in each batch and the batch selection scheme (sparse vs. compact) are tunable parameters in reconstruction. Here we use the sparse selection. In addition, the number of probe modes in mixed-states ptychography(*48*) can be adjusted to account for partial coherence and probe vibration(*49*) . In the orthogonal probe relaxation (OPR) technique(*50*) , which is often used to reduce artifacts caused by probe variation in a single scan, the number of orthogonal modes kept in truncated singular value decomposition controls the number of structural changes allowed at each scan position. The regularization algorithm(*51*, *52*) is tested, while it will suppress the separation of individual layers.

Therefore, it is not used in the reconstruction. Moreover, position correction can refine inaccurate, scan positions and intensity correction accounts for changes in probe intensity. Lastly, the "multimodal" option specifies if all or only the first probe mode is used to update the object function. The detailed comparison of different parameters and other reconstruction options are discussed in **Supplementary Information Note.** After ptychography reconstruction**,** the atom positions on each layer are extracted by the CalAtom package(*53*) .

Nano-beam 4D-STEM datasets (**Fig. S1**), utilizing a 0.5 mrad convergence angle, were employed to ascertain lattice information across the sample. A camera length of 720 mm was used to ensure a clear delineation of the twelve first-order diffraction spots. Analysis of the center of mass of these spots facilitated the mapping of lattice strain, layer rotation, and the twist angle between two layers with sub-picometer precision in a large field of view (approximately 300 nm).

**4D-STEM Simulation**

The simulated datasets in **Fig. S4**, **Fig. S5** are generated from the abTEM package(*54*) through multislice simulation with the similar parameters with the experimental data, except the dose series. Then the reconstructed images are generated through the MEP method with the same reconstruction parameters, except the number of probe modes. Since the simulated probe has less partial coherence, only two probe modes are used.

**Acknowledgements**

The work is supported by National Science Foundation CMMI-2239545 and the Welch Foundation (C-2065). C.S. and Y.H. acknowledge the Electron Microscopy Center, Rice University. Y. J. acknowledges support from the U.S. DOE Office of Science-Basic Energy Sciences, under Contract No. DEAC02-06CH11357. Y.L., Y.Luo., G.E., M.H.N. and C.K.S. acknowledge the support by the NSF through the Center for Dynamics and Control of Materials: an NSF Materials Research Science and Engineering Center (Cooperative Agreement No. DMR-2308817) and NSF DMR-2219610. G.E. and M.H.N. acknowledge the Texas Advanced Computing Center (TACC) at The University of Texas at Austin for providing computational resources that have contributed to the research results reported within this paper. This work also used computational resources from Stampede3 at The University of Texas at Austin through allocation PHY250206 from the Advanced Cyberinfrastructure Coordination Ecosystem: Services and Support (ACCESS) program, which is supported by National Science Foundation grants 2138259, 2138286,~2138307, ~2137603, and ~2138296.